\documentclass[prl,aps,twocolumn,showpacs]{revtex4}

\usepackage{epsfig}
\usepackage{amsmath}

\begin{document}

\title
{
Nuclear spin relaxation probed by a single quantum dot 
}

\author{A.~K.~H\"uttel$^1$}
\author{J.~Weber$^1$}
\author{A.~W.~Holleitner$^{1,2}$}
\author{D.~Weinmann$^3$}
\author{K.~Eberl$^4$}
\author{R.~H.~Blick$^{1,5}$} 

\affiliation{$^1$Center for NanoScience and Sektion Physik, 
Ludwig--Maximilians--Universit\"at, Geschwister--Scholl--Platz 1, 
80539~M\"unchen, Germany}
\affiliation{$^2$California NanoSystems Institute (CNSI), 4670 Physical Sciences
Building North, University of California, Santa Barbara, CA 93106-6105, USA.}
\affiliation{$^3$Institut de Physique et Chimie des Mat\'{e}riaux de
  Strasbourg, UMR 7504 (CNRS-ULP), 
  23 rue du Loess,
  67037 Strasbourg, France}
\affiliation{$^4$Max-Planck-Institut f\"ur Festk\"orperforschung, 
Heisenbergstra{\ss}e 1, 70569 Stuttgart, Germany}
\affiliation{$^5$Dept. of Electrical \& Computer Engineering, University of Wisconsin
 -- Madison, 1415 Engineering Drive, Madison, WI 53706-1691, USA.}

\date{June 11, 2003}

\begin{abstract}
We present measurements on nuclear spin relaxation probed by a single
quantum dot in a high-mobility electron gas. Current passing through the dot leads to a spin transfer
from the electronic to the nuclear spin system. Applying electron spin resonance the
transfer mechanism can directly be tuned. Additionally, the dependence of
nuclear spin relaxation on the dot gate voltage is observed. We find
electron-nuclear relaxation times of the order of 10 minutes. 
\end{abstract}

\pacs{
73.21.La,    
72.25.Rb,    
76.60.Es     
}
\maketitle

{\it Introduction}-- The ever increasing demand for computing power as well as
theoretical considerations on 
the basic notions of information processing~\cite{info_is_physical}
have led to the development of the new concept of quantum
computing~\cite{textbook_on_computing}. Different experimental systems have
been suggested performing quantum computational 
tasks~\cite{bouwmeester_poqi}. 
Among the most promising of these are quantum dots~\cite{review} which can by now be
fabricated with great accuracy  
in a whole variety of circuits enabling not only probing molecular binding mechanisms in coupled 
dots~\cite{science_holleitner}, but also the definition of quantum
bits~\cite{pra_loss,ieee_blick}.  

One of the key questions in quantum information processing is how to efficiently store such quantum
bits with a sufficient life time. As suggested by Kane~\cite{kane} one system for achieving 
this would be a tunable electron-nuclear spin system, such as a quantum dot coupling to nuclear 
spins of the embedding crystal matrix. 
For isolated electron spins trapped in electrostatically defined quantum dots,
the theoretical possibilities of realizing qubit operations have
already been investigated in great detail~\cite{pra_loss}. Furthermore,
it has been demonstrated that controlled spin transfer between electrons and
nuclei is possible in spin polarized two-dimensional~\cite{nature_smet} 
and one-dimensional systems~\cite{prl_wald} and can be detected using electron or nuclear 
spin resonance techniques~\cite{prl_dobers_88}. 

In constrast to these earlier works which beautifully demonstrated tuning of the coupling of
a two-dimensional electron system (2DES)
to the nuclear spin lattice, as well as controlling and manipulating nuclear spin
relaxation~\cite{nature_smet}, we focus on the interaction of 
electrons confined in a single quantum dot with a much smaller number of
nuclei, approaching a mesoscopic regime and strongly localizing the polarization.
In addition, we address recent work by Lyanda-Geller~{\it et al.}~\cite{lyanda} who consider 
nuclear spin relaxation (NSR) caused by a quantum dot coupling to the
nuclear magnetic moments. The nuclear system's relaxation time can be
several hours, being perfectly suited for phase coherent storage
of quantum information.
As will be seen below we find according to the predictions of Ref.~\cite{lyanda} 
the relaxation time to depend on the single electron tunneling resonance
condition of the dot. 
We have to note that in our experiment we concentrate  
on a single quantum dot as compared to recently reported measurements by 
Ono and Tarucha~\cite{ono02:H245} on a coupled dot.

The main ingredient of our approach is the preparation of a specific quantum
dot state, adjacent to a region of spin blockade of transport as discussed 
earlier~\cite{rok,weinmann,andi}. As our measurements in a bi-axial magnet at low fields show,  
the dot state also possesses a large angular momentum $L$. This 
effect leads to a spin current, continuously flipping nuclear spins and hence transferring
and storing the magnetic momentum through the hyperfine 
interaction. As expected for this case, electron spin resonance strongly enhances the
NSR. This part of the experiment is based on earlier work on photon assisted
tunneling in quantum dots~\cite{leo,robert,robert_esr} and is partly inspired
by theoretical work of Engel and Loss~\cite{engel}. 

\begin{figure}[tb]
\begin{center}
\epsfig{file=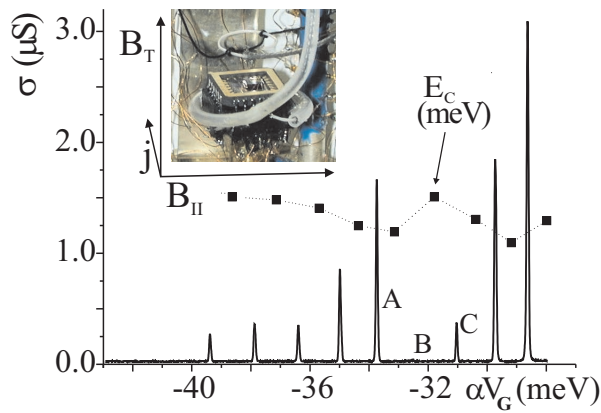, width=8cm}
\end{center}
\caption{
Coulomb blockade oscillations of conductance $\sigma = dI/dV$ of a single
quantum dot vs. gate voltage $V_G$. 
The dotted line gives the charging energy $E_C$ for adding single electrons.
As seen the energy assumes a local maximum between peak B and C.
Inset: sample holder setup including radiowave and microwave antennae. 
The AC signal in the antennae leads to an alternating magnetic field perpendicular to the 
sample's surface. 
}
\label{fig1}
\end{figure}

{\it Methods}-- A typical conductance trace characterizing the quantum dot 
is shown in Fig.~\ref{fig1}. The dot measured here is defined electrostatically in the 
2DES of an epitaxially grown AlGaAs/GaAs
heterostructure: a split gate geometry is written by electron beam
lithography on the crystal surface. By negatively charging the gate electrodes, in the 2DES
$120\,\mathrm{nm}$ below the surface a quantum dot containing approximately $85$ electrons is formed. The data are taken at
a bath temperature of $40\,\mathrm{mK}$ and an electron temperature of
$\sim 80\,\mathrm{mK}$
in a $^{3}$He/$^{4}$He dilution refrigerator system.
A similar conductance pattern as in Fig.~\ref{fig1} was obtained in our earlier work on spin blockade in
a dot containing about 50 electrons~\cite{andi}. 
At $4.2\,\mathrm{K}$ the carrier density of the 2DES is $1.8
\times 10^{15}\,\mathrm{m}^{-2}$~and the electron mobility is
$75\,\mathrm{m}^{2}/\mathrm{Vs}$. For DC measurements, a source-drain voltage of $20\,\mu$V
is provided. The addition energy of the dot is given by $E_C \sim 1.25\,\mathrm{meV}$, as indicated
by the dotted line in Fig.~\ref{fig1}. 

\begin{figure}[tb]
\begin{center}
\epsfig{file=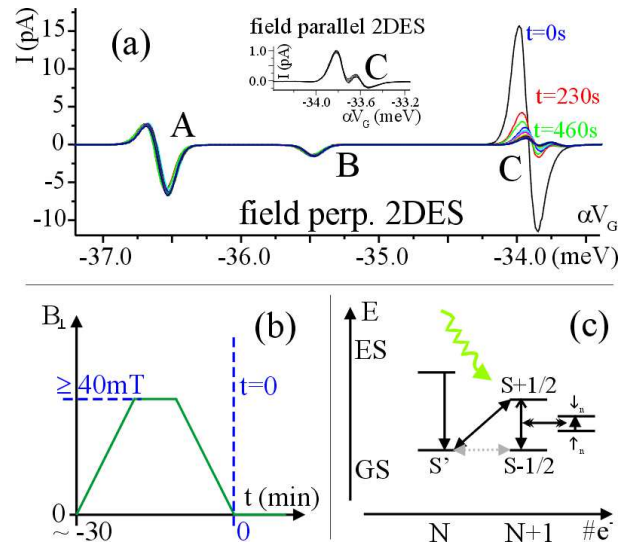, width=8cm}
\end{center}
\caption{(a) Quantum dot photocurrent peaks A,~B, and C under microwave
radiation at 10.01~GHz. Prior to taking these data the perpendicular magnetic
field was ramped from $B_{\perp}$ = 0~T to 487~mT in $t_r \sim $11~min,
maintained at this value for $t_{m} \sim 8$~min, and reduced to $B_{\perp}$ =
0~T within 11 min. A strong memory effect at peak C can be observed. 
The inset gives the same measurement for peak~C, using a parallel field
orientation. Obviously no long-term memory is found.
(b) Schematic plot detailing $B(t)$ in the measurement setup. 
(c) Level diagram for the transition from $N$ 
to $(N+1)$ electrons at peak C (see text for further details). 
}
\label{fig2}
\end{figure}

{\it Experiment}-- In the conductance spectrum of Fig.~\ref{fig1}, a sequence of three peaks
is marked by the letters A, B, C. Peak A displays conventional conductance, whereas
peak B is nearly completely blocked at low transport voltage and peak C
shows a response smaller than average. In subsequent measurements, 
the suspended loop antenna visible in the inset of Fig.~\ref{fig1} is  
emitting microwave radiation onto the sample chip: Fig.~\ref{fig2}(a) again displays
the three peaks now showing the induced photocurrent under irradiation at 10.01~GHz. 
It is important to note that prior to taking these data traces the perpendicular magnetic
field was ramped from $B_{\perp}$ = $0\,\mathrm{T}$ up to $B_{\perp,\mathrm{max}} \approx 0.5\,\mathrm{T}$
in $t_r= 11\,\mathrm{min}$, maintained at this value for $t_{m} =
8\,\mathrm{min}$, and subsequently reduced to
$B_{\perp} = 0\,\mathrm{T}$ within $11\,\mathrm{min}$.
As seen peak A gives the conventional rectification signal with a forward and backward pumped 
current~\cite{weis}. Surprisingly, the spin blockade transition at~B reveals a
backward current only. This can be explained by strongly differing excitation
energies of the quantum dot at subsequent electron numbers, consistent with level
scrambling causing spin blockade type-II. 

The assumption of a certain spin texture gains evidence when focussing on
resonance C -- which, being located next to the spin blocked peak, can feature a high spin as well: 
after ramping $B_{\perp}$ the relaxation of the current trace 
requires additionally more than $10\,\mathrm{min}$. Testing the available parameter ranges,  
we found $B_{\perp,\mathrm{max}} = 40\,\mathrm{mT}$, as well as ramp times and a waiting period of
$t_{r/m}=6\,\mathrm{min}$ to be sufficient for clearly demonstrating the effect. 

This time dependence is attributed to the observation of a slowly decaying
nuclear spin polarization, which has been induced during the magnetic field sweep by dynamic
polarization processes. Here, the accessibility of transport channels depends
on the population of spin states~\cite{nature_smet,prl_wald}.
In a quantum dot in an AlGaAs/GaAs-heterostructure 
with a diameter and height of 125~nm and 10~nm $\tilde{N} \sim 2.2 \times
10^7$ nuclei are engulfed by the electronic volume. For comparison, a rough
estimate gives $10^{8}$ -- $10^9$ electrons passing the dot during the magnetic field
sweep of $30\,\mathrm{min}$. In addition, a completely polarized nuclear
spin population has been shown to give rise to local magnetic fields of up to several
Tesla~\cite{nature_smet, paget}. Even partial polarization or polarization within a
small volume is expected to have a clearly visible effect. 

The inset in Fig.~\ref{fig2}(a) gives an identical measurement for peak~C in a
cycled magnetic field parallel to the 2DES
-- as seen no memory effect
is observed. This leads us to the conclusion that orbital effects
bound to a particular spin state
are responsible for coupling to the nuclear magnetic moments. 
A pure spin flip would obey Zeeman splitting 
in a parallel magnetic field as well, and the phenomenon should persist in
this case. In an intuitive picture, at peak~C the electrons tunneling through the quantum dot
can be thought to be passing through a high-$L$ state, circulating at the 
edge and allowing to transfer momentum from the electronic to the nuclear system.

A possible level scenario of the spin flip operation is given in the diagram
of Fig.~\ref{fig2}(c): as measured, we assume the direct transition probability between the $N$
and $(N+1)$ electron ground states to be low; single electron tunneling is
partly suppressed.
An increase in current via the excited $(N+1)$ electron state
takes place as soon as irradiation enhances the energy available. 
Relaxation into the ground state via hyperfine coupling to the
nuclear spin system comprises a change in  
spin quantum number by $\Delta S = 1$, spin conservation in the
hyperfine interaction results in a flip-flop process of electron and nuclear spins~\cite{prl_wald}. 
This brings the spin of a nearby nucleus from $\left|\downarrow_{\mathrm{n}}\right>$ into the 
state $\left|\uparrow_{\mathrm{n}}\right>$. The dot remains in the $(N+1)$
electron ground state until the electron tunnels out via the
ground state transition and the cycle restarts. On the timescale given, the spin flip
rate required for polarization is consistent with theoretical predictions for
a similar quantum dot~\cite{erlingsson}, where the energy mismatch between
electronic and nuclear Zeeman splitting, otherwise suppressing this process,
is compensated by phonon emission.

However, the relaxation by flip-flop processes is only possible as long as sufficient
nuclei with appropriate spin direction are available. Assuming a nonzero
polarization, hyperfine relaxation decreases, and the trapping effect
described above is deactivated, leading to an increase in current. This gives
a possible mechanism of detection of the gradual depolarization after ramping
down the magnetic field. Other mechanisms include a shift of electronic
levels induced by a remaining effective nuclear magnetic field.
The quantum dot operates as partial spin filter and inverter; a weak
polarization of the nuclear spins even without supporting microwave 
radiation is possible, as long as a magnetic field perpendicular to the
surface provides an orientation.

Subsequently we want to address the change in nuclear relaxation time in dependence of
the quantum dot's resonance state 
as Lyanda-Geller~{\it et al.} investigated in their calculations~\cite{lyanda}: 
Again we focus on resonance~C in a 
perpendicular field orientation with the field cycling as introduced above. 
\begin{figure}[tb]
\begin{center}
\epsfig{file=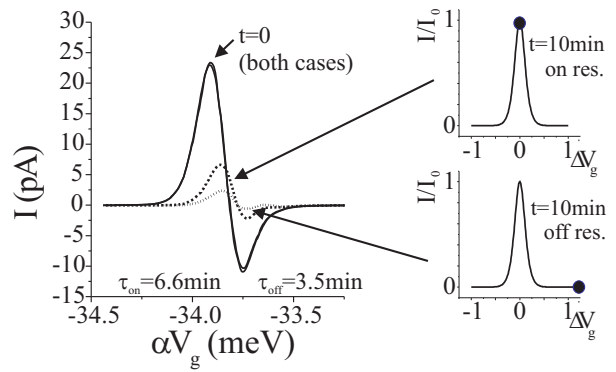, width=8cm}
\end{center}
\caption{
Dependence of NSR on Coulomb blockade: after cycling 
$B_{\perp}$ as described in the text, and taking a trace for reference at $t=0$ (solid
lines) the gate voltage is kept fixed for a waiting period of $\Delta t = 10\,\mathrm{min}$.
Relaxation is then determined in dependence of the gate voltage position.
Relaxation times strongly vary comparing waiting positions in resonance and off resonance.
}
\label{fig4}
\end{figure}
The main difference now is that
relaxation of the photocurrent trace after switching off $B_{\perp}$ is not
monitored sweeping continuously over the gate voltage range. A first current
trace is recorded; then the gate voltage is kept either at SET
resonance $V_g^{\rm{res}}$ or off resonance $V_g^{\rm{off}}$, as
shown schematically in the insets of Fig.~\ref{fig4}. 
$10\,\mathrm{min}$ later, an additional trace of the peak is taken.  
Obviously, in the case of SET resonance the relaxation slows down
considerably. As shown by the authors of Ref.~\cite{lyanda}, a non-negligible
spin-orbit interaction~\cite{andi} in combination with the differing nature of coupling
processes in separate gate voltage regimes causes such behaviour.

In analyzing the relaxation process quantitatively we compare the integrated difference of
relaxed and excited photocurrent traces and finally normalize it with respect to the relaxed curve.
This is defined by a function
\begin{equation} \label{Auslenkung}
E[t] \equiv
 \frac{\int\limits_{V_{G1}}^{V_{G2}}dV_{G}|I(V_{G},t)-I(V_{G},\infty)|}
            {\int\limits_{V_{G1}}^{V_{G2}}dV_{G}|I(V_{G},\infty)|}.
\end{equation}
The characteristic decay time constant of $E[t]$ corresponds to the
nuclear spin relaxation 
time $T_1$ and typically assumes values of $\tau \sim 5 \dots 12\,\mathrm{min}$.
In the measurement described above, we find as relaxation times for $E[t]$ depending on the gate
voltage during waiting $\tau_{\mathrm{res}} = 6.6$~min and 
$\tau_{\mathrm{off}} = 3.5$~min, hence again supporting the theoretical
assumptions of Lyanda-Geller {\it et al.}

\begin{figure}[tb]
\begin{center}
\epsfig{file=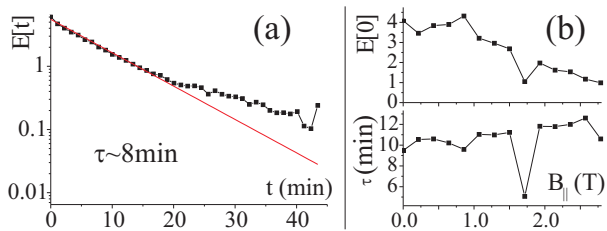, width=8cm}
\end{center}
\caption{
(a) Exemplary plot of the relaxation of peak C as seen in
Fig.~\ref{fig2}(a) with a relaxation constant 
of $T_1 \sim \tau \sim 8\,\mathrm{min}$. The function $E[t]$ as a measure for
the peak relaxation and thereby the nuclear spin polarization is defined in the
text.
(b) Electron spin resonance (ESR) in an additional constant parallel magnetic
field found by comparing the maximal polarization 
$E[t=0]$ and the relaxation times $\tau$. Mixing of ground and excited
states leads to a strongly reduced $\tau$ and $E[0]$. 
}
\label{fig3}
\end{figure}

In extending the discussion above we now can apply classical electron spin resonance to tune the 
nuclear relaxation time. This is performed by again irradiating at
$10.01\,\mathrm{GHz}$ and measuring $E[t]$ for different values of an additional parallel
magnetic field which couples to the spin only. The perpendicular field is
sequentially polarizing the nuclear  
spins through the quantum dot at 200~mT. The maximal 
amplitude at $t=0$, i.e. directly after the external field $B_\perp$ has been
brought to zero, is given by $E[0]$ and corresponds to maximal polarization
of the nuclear spin system within our interval of observation. $E[t]$ then
decays exponentially, as can be seen in the exemplary plot of Fig.~\ref{fig3}(a).
In Fig.~\ref{fig3}(b) at 10.01~GHz we obviously find for a field of
$B_{\parallel} = 1.7$~T a clear resonant feature in the NSR time which agrees
with the value obtained from $E_Z = g \mu_B B$ assuming $g=-0.42$. Particularly, the life time
reduction in resonance supports the level diagram sketched in
Fig.~\ref{fig2}(c): the electron spin resonance leads to a mixing of the
ground and excited states with a spin change of $\Delta S = 1$. This it to be
considered as bypassing the pumping of nuclear spins through an electron spin transition. As seen we
are able to achieve a change of over 50\% in NSR time.


{\it Conclusions}-- Relating to SET blockade regimes in a single quantum dot we find 
strong coupling of electron and nuclear spins via the hyperfine
interaction. This leads to measured nuclear relaxation times exceeding 10~min. 
In accordance with reference~\cite{lyanda} we find that NSR is maximal in the
regime of Coulomb blockade. Electron spin resonance is applied to broadly vary
NSR. 
As we observe electron-nuclear spin coupling at moderate fields of some 50~mT in conjunction with 
the tuning mechanisms introduced, we conclude that -- although in our case
still a large
number of nuclei is addressed simultaneously -- this will strongly support quantum information 
processing in solid state systems, being a first step towards quantum state
transfer and the long-term storage of quantum spin information.

{\it Acknowledgements}-- We like to thank J.P.~Kotthaus, R.~Gross, and 
R.~Jalabert for detailed discussions and A.~Kriele and K.~Neumaier for expert technical help.
We acknowledge financial support by the Deutsche Forschungs\-ge\-mein\-schaft
through the Schwerpunkt 'Quan\-ten\-in\-for\-ma\-tions\-ver\-ar\-bei\-tung' (Bl/487-2-2)
and the Defense Advanced Research Projects Agency (EOARD project: F61775-01-WE037).
AKH gratefully acknowledges support by the Studienstiftung des deutschen Volkes.

\bibliographystyle{apsrev}

\end{document}